Disaster Recovery Using Virtual Machines

**Dr. Timur Mirzoev**
Department of Information Technology
College of Information Technology
Georgia Southern University

## Introduction

Today, the importance of having 100% uptime for businesses and industries is clear –

financial reasons and often strict government regulations for certain industries require 100%

business continuity. The concept of business continuity (BC), as Microsoft defines it - "the

ability of an organization to continue to function even after a disastrous event, accomplished

through the deployment of redundant hardware and software, the use of fault tolerant systems, as

well as a solid backup and recovery strategy" (Microsoft 2005) directly relates to an

organization's ability to quickly restore and deploy IT backups and business operations in a short

period of time. Disaster recovery's precedence determines the quality of restored business

operations and allows for continuation of business. Nowadays, backups and server uptime are a

must; however, those are not directly related and not sufficient enough to provide grounds for

disaster recovery for a company. The advancement of information technologies to the level of

virtual environments suggests easier deployment and restoration of IT functions once

virtualization is utilized for BC management. Evident cost reduction and elimination of physical

hardware components along with space savings are provided with virtualization processes

(Goldworm, B., Skamarock, A. 2007); those serve as vital components for any disaster recovery process.

This manuscript presents several necessary steps to be taken before and after a creation of a disaster recovery plan. The amount of resources, administration/personnel involvement and financial costs associated with a design of a Business Continuity Plan (BCP[1]), simply do not allow for a quick and easy BCP implementation. Along with the procedures associated with a BCP design, this study suggests several concepts for BCP and BC utilizing virtualization software which allows for a significant cost and time reduction in preparation for a successful disaster recovery plan. Furthermore, it is important to understand that disaster recovery planning is a separate process from business continuity management.

---

[1] Business Continuity Institute defines Business Continuity Planning as Business Continuity Management (BCM).

## Business Continuity Management Cycle

Preparation for something unknown or unexpected is a difficult task. It is a challenge for organizations to embrace every business operation and function along with the required IT support and functionality and then downsizing them to 2-3 mobile office/recovery units. Mobile office units could be contracted out via disaster recovery agreements by companies such as Hewlett-Packard, Centurion, Agility Recovery Solutions and many others. An example of a mobile office unit is presented by *Figure 1*. It is also important to realize that BC process should incorporate *every* vital business function that allows for continuity and requires extensive preparation before signing any contracts. *Figure 2* presents the Business Continuity Management Cycle which provides initial and at the same time continuous steps in BC management. The Cycle starts with assessing risks and business impacts once various threats and levels of a possible disaster are understood. Once risk assessment takes place then it is possible to realize the minimal accepted business functionality for various disaster recovery scenarios. The presented cycle is a continuous process that allows for assessment, planning, creation, testing and improvement of a business continuity plan.

- Nationwide Delivery in 24 to 48 Hours
- Deployable within 3 Hours
- On-Board Diesel Generator
- Pre-Installed Voice & Data CAT-5 Cabling
- HVAC

- Internal Alarm System
- Network Connectivity to Additional Units
- Office Furniture
- Self Leveling Capabilities
- No Special Permits or Travel Restrictions

**Recovery Configurations**

- Call Center
- Command Center
- Mobile Bank
- Data Center
- End User Workspace
- Executive Offices

**Options**

- Satellite
- Conference Room
- Break Room
- Voice Communications
- On-Site Support
- Multi-Site Coverage on Single Contract
- Restrooms
- Water

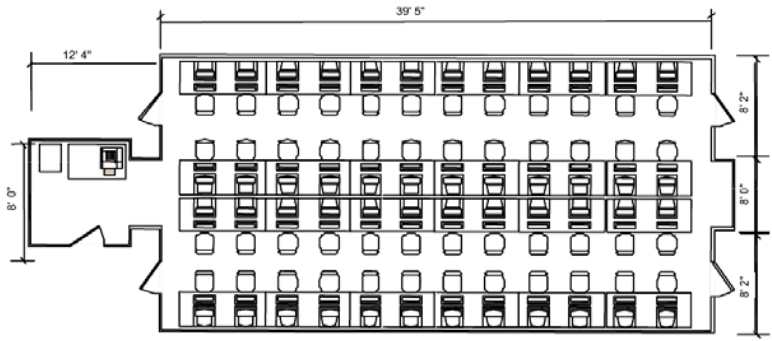

*Figure 1.* Mobile recovery unit.

Source: http://www.centuriondr.com/

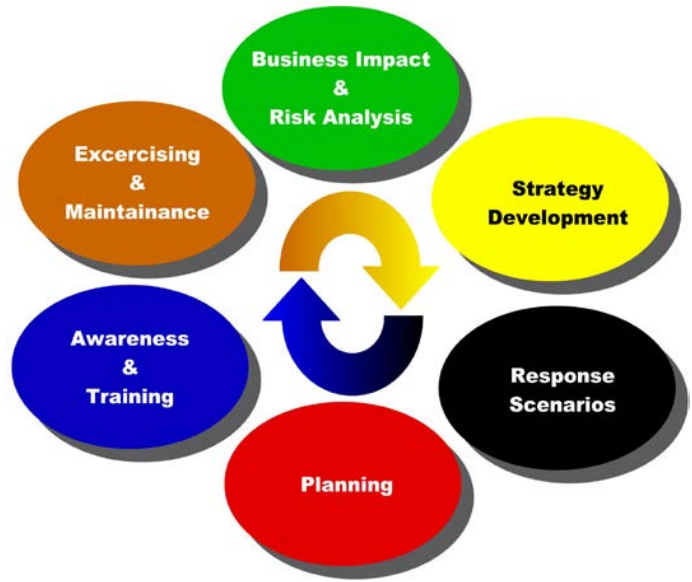

*Figure 2.* Business Continuity Management Cycle.

Prior to taking any steps towards disaster recovery planning, the following questions need to be answered: 1) does a company's administration approve and support the disaster recovery planning? 2) what personnel is going to organize, manage and create the plan? 3) which essential functions need to be restored? 4) what are the federal/state government guidelines for this particular industry? Those initial questions about BCP would be addressed by a Business Continuity Committee (BCC) that should be formed as soon as the executive/administrative approval and support for BCP is obtained. Strategy development should provide detailed steps for each member of an Emergency Response Team (ERT) which should be formed by BCC.

### Planning and Implementation

The following essential steps in disaster recovery could serve as a guide to BC planning; each of these steps incorporates several other processes which require a high level of commitment, time, money and organizational discipline (MIT 1995):

1. **Obtain Top Management Commitment.** Top management must support and be involved in the development of the disaster recovery planning process. Management should be responsible for coordinating the disaster recovery plan and ensuring its effectiveness within the organization.

2. **Establish a planning committee.** A planning committee should be appointed to oversee the development and implementation of the plan. The planning committee should include representatives from all functional areas of the organization.

3. **Perform a risk assessment.** The planning committee should prepare a risk analysis and business impact analysis that includes a range of possible disasters, including natural, technical and human threats.

4. **Establish priorities for processing and operations.** The critical needs of each

department within the organization should be carefully evaluated in such areas as:

- o Functional operations
- o Key personnel
- o Information
- o Processing Systems
- o Service
- o Documentation
- o Vital records
- o Policies and procedures

5. **Determine Applicable Recovery Strategies.** Determine the most practical alternatives

for processing in case of a disaster should be researched and evaluated:

- o Facilities
- o Hardware
- o Software
- o Communications
- o Data files
- o Customer services
- o User operations
- o MIS
- o End-user systems

6. **Perform data collection.** This step is the most time consuming.

- o Backup position listing
- o Critical telephone numbers
- o Communications Inventory
- o Distribution register
- o Documentation inventory
- o Equipment inventory
- o Forms inventory
- o Insurance Policy inventory
- o Main computer hardware inventory
- o Master call list
- o Master vendor list
- o Microcomputer hardware and   software inventory
- o Notification checklist
- o Office supply inventory
- o Off-site storage location inventory
- o Software and data files backup/retention schedules
- o Telephone inventory
- o Temporary location specifications
- o Other materials and documentation

7. **Organize and document a written plan.** An outline of the plan's contents should be

prepared to guide the development of the detailed procedures. The structure of the contingency organization may not be the same as the existing organization chart.

8. **Develop testing criteria and procedures.** It is essential that the plan be thoroughly tested and evaluated on a regular basis (at least annually). Procedures to test the plan should be documented.

9. **Test the Plan.** After testing procedures have been completed, an initial test of the plan should be performed by conducting a structured walk-through test. Types of tests include:

   o Checklist tests
   o Simulation tests
   o Parallel tests
   o Full interruption tests

10. **Approve the plan.** Once the disaster recovery plan has been written and tested, the plan should be approved by top management.

The steps in BC management presented in this chapter embrace many business functions and operations. Today, each of these operations is not functional without an extensive IT infrastructure; furthermore, industries such as healthcare endure strict government regulations for IT and business functions which further complicate disaster recovery process. If at the time of initial phases of BCP organization does not operate in a virtual IT environment it is a perfect time to utilize virtualization in preparation of a disaster recovery plan. There are many courses and even schools that provide education and certificates that allow organizations and industries to apply BCM knowledge towards disaster recovery planning.

*Virtualization Applied to BCM*

Virtual technologies have a broad range of contexts – operating systems (OS), programming languages, computer architecture (Nair, R., Smith, J. 2005). Virtualization of operating systems and computer architectures significantly benefits any disaster recovery process and would improve business continuity. Depending on which OS environment is used, a company should provide backup of virtual workstations, servers instead of the conventional images of computers. As it was mentioned earlier the conversion of computer systems into virtual machines at the time of BCP creates an opportunity for a company to save money, resources and time. Tools such as VMware Converter allow conversions of various computer platforms into VMware virtual machines. For example, instead of providing images of several servers to a company that deploys mobile office units, an organization should provide 1 image which contains all virtual servers. *Figure 3* provides an example of how that conversion is performed. Conversion to virtual servers would allow reduction of costs by contracting out fewer PCs identified in a disaster recovery agreement.

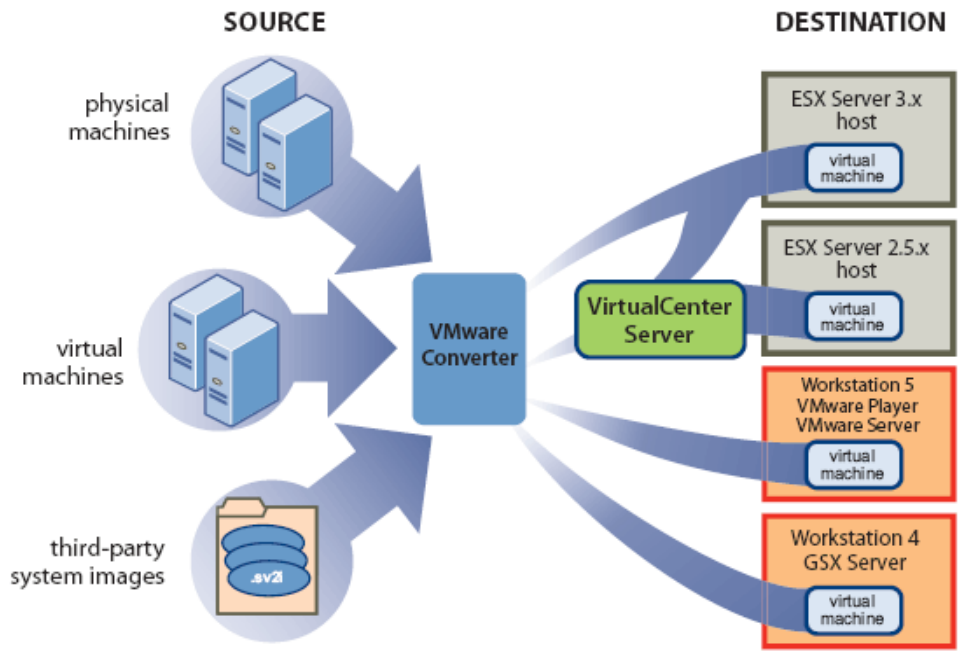

*Figure 3.* Performing a conversion of a physical machine into a virtual one.

Source: http://www.vmware.com/products/converter/

Virtualization concept is not a new idea. It is based on a time-sharing concept originally developed by scientists at Massachusetts Institute of Technology (MIT) in 1961 (Corbato, F. *et al.* 1961). Time-sharing creates an opportunity of concurrently managing multi-hardware and multiuser environments on a single physical machine. Today, many vendors such as IBM, Sun Microsystems, HP and others have taken this time-sharing concept further and developed virtualization scheme of various types including Integrity VM by HP (Herington, D., Jacquot, B. 2006). The advantages of modern technologies such as Integrity VM allow running any operating system inside VM that supports Integrity VM platform (Herington, D., Jacquot, B. 2006). An example of how virtualization is applied in a hosted virtualization model is shown by *Figure 4* (Goldworm, B., Skamarock, A. 2007).

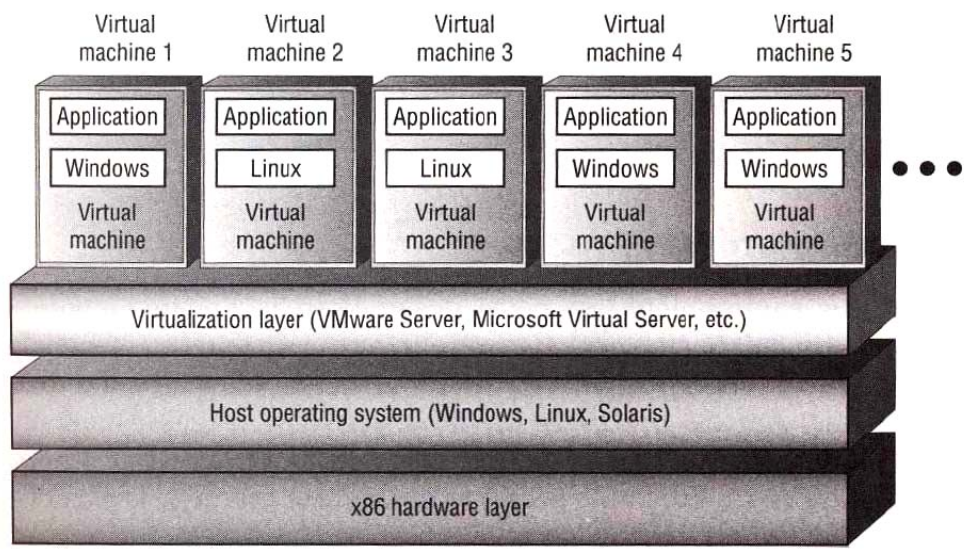

*Figure 4*. Hosted approach for virtual servers

Source: Goldworm, B., Skamarock, A. (2007). Blade Servers and virtualization. Transforming Enterprise Computing while Cutting Costs. p.97

A step further in disaster recovery using virtual technologies could be taken with virtual network environment and storage virtualization. Storage virtualization allows for separation of physical management from the application software on a server which, once again cut the costs of create backup sties and backup hardware in the event of a disaster (Goldworm, B., Skamarock, A. 2007).

## Conclusion

As information technologies progress continuously, it is important to realize practical applications of available IT resources and direct them towards every day operations. For instance, companies should fully deploy virtualization for disaster recovery and business continuity even if regular business operations do not require virtualization. Reduction of costs, time and resources is going to have a positive impact on business continuity management.